\documentclass[12pt]{article}
\usepackage{amssymb}
\usepackage{amsfonts}
\usepackage{graphicx}

\newcommand{\real}{\mathbb{R}}

\newtheorem{thm}{Theorem}[section]
\newtheorem{remark}{Remark}[section]

\title{The Born Oppenheimer wave function near level crossing}
\author{J.~E.~Avron and A. Gordon
\\ Department of Physics, Technion, 32000 Haifa, Israel}
\begin{document}
\maketitle
\begin{abstract}

The standard Born Oppenheimer theory does not give an accurate
description of the wave function near points of level crossing. We
give  such a description near an isotropic conic crossing, for
energies close to the crossing energy. This leads to the study of
two coupled second order ordinary differential equations whose
solution is described in terms of the generalized hypergeometric
functions of the kind $_0F_3(;a,b,c;z)$. We find that, at low
angular momenta, the mixing due to crossing is surprisingly large,
scaling like $\mu^{1/6}$, where $\mu$ is the electron to nuclear
mass ratio.
\end{abstract}

\section{Introduction}

In 1927, in a landmark paper, Born and Oppenheimer \cite{bo} paved
the way to applying quantum mechanics to molecular spectra. In
their paper they introduced an approximation that greatly
simplified the treatment of quantum mechanical spectral problems
in which the particles can be divided into heavy and light.
Molecules are an example since the nuclei are much heavier than
the electrons. We shall denote by $\mu$ the small parameter of the
theory. In molecules $\mu\sim 10^{-4} $, the electron to nucleon
mass ratio. Since the light particles are associated with the fast
degrees of freedom and the heavy particles with the slow degrees
of freedom, the Born Oppenheimer approximation is related to the
adiabatic approximation \cite{elgart,borneman,kato-ad}. At the
same time, the Born Oppenheimer method can be viewed as a
generalized semiclassical approximation where the small parameter
$\mu$ plays the role of $\hbar^2$. This is a reflection of the
fact that the electrons and nuclei also live on different spatial
scales, the electronic wave function is spread and is far from
the semiclassical limit, while the nuclear wave function is tight
and close to semiclassical.

The procedure put forward by Born and Oppenheimer is to first
solve the electronic spectral problem with fixed nuclei, and view
the nuclear coordinates as parameters.  To the leading order in
$\mu$, and far from crossings, the heavy degrees of freedom in
Born Oppenheimer theory are described by a (scalar) Schr\"odinger
operator in the semiclassical limit with $\hbar^2=\mu$ and where
the electronic energy surface serves as a potential.

 Born Oppenheimer theory developed into two distinct directions.
The main direction has been the application to various systems and
the development of effective and accurate methods of calculations
\cite{mead,mead2,mead_truhlar,jackiw}. The second direction has
been the development of the theory as a tool of rigorous spectral
theory \cite{aventini-seiler,borneman,hagedorn} . Our work falls
into the first class.

Points where electronic energy surfaces cross are singular points
of the Born Oppenheimer theory. In certain cases, these points can
affect spectral properties  \cite{mead}. In this work we focus on
the behaviour of eigenfunctions near crossing. There is surprising
little that is known about this. It is not known if the function
has finite values at the crossing; how the amplitude of the wave
function near the crossing scales with $\mu$. Since crossing
controls the mixing of electronic levels, the knowledge of the
wave function near crossing is important. Of course, these
questions are interesting in the case that the crossing lies in
the classically allowed region.

In this work we address these issues for conical (i.e. linear)
crossing of two levels where the Born Oppenheimer problem reduces
to two coupled Schr\"odinger equations. In the isotropic case the
analysis reduces further to the study of two coupled, second
order, ordinary differential equations. We obtain the nuclear wave
function analytically, to leading order in $\mu$, close to the
crossing. It is related to the generalized hypergeometric
functions of the kind $_0F_3$. This function takes the ordinary
Born Oppenheimer nuclear wave functions, which are a good
approximation far from the crossing, all the way to the crossing,
where the nuclear wave function mixes the two electronic levels.
We find that the nuclear wave function with total angular
momentum\footnote{We do not actually mean here the physical
angular momentum, but a quantum number reminiscent to it, see Eq.
(\ref{jz}).} $m=\pm \frac 1 2$ is {\it nonzero} at the crossing
point. Moreover, for low momenta, the wave function has a large
amplitude near the crossing, of order $\mu^{-1/4}$. We find
appreciable mixing of the two levels at distances that are smaller
than $\mu^{1/3}$ and the total weight that is mixed between levels
scales like $\mu^{1/6}$ which is remarkably large.

\section{The Born Oppenheimer approximation}
This section  is a brief introduction to the basic and elementary
elements of Born Oppenheimer theory.
\subsection{The basic model}
A prototype of the Born Oppenheimer problem, and the one we study
here is \cite{mead}:
\begin{equation}\label{generalBO}
H=-\mu \Delta_x + H_e(x)
\end{equation}
where $\mu$ is a small parameter. $H_e(x)$ is an operator valued
function that acts in the Hilbert space of the light degrees of
freedom. Since the eigenvalues of $H_e(x)$ repel \cite{wvn}, one
does not expect crossing in the case of one heavy coordinate $x$.
If $H_e(x)$ is time reversal invariant, then stable crossing will
occur if there are two heavy coordinates. We therefore assume that
$x\in \mathbb{R}^{2}$. For the sake of simplicity we have taken
identical masses for the two heavy degrees of
freedom\footnote{There is no loss here, for by scaling some of the
$x$-directions this can always be achieved}. Two degrees of
freedom
 is the simplest case that is still rich enough to
cover the phenomena we are interested in.

\begin{figure}[htb]
  \centering
  \includegraphics[height=4.cm]{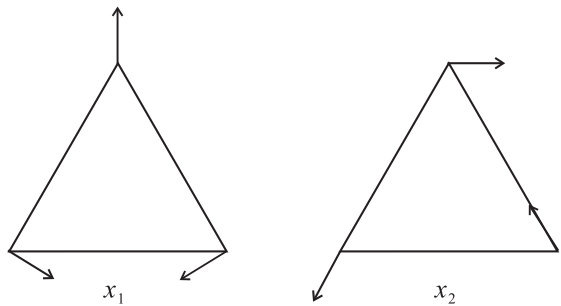}
  \caption{\small Two degenerate modes in triangular molecules.}\label{modes}
\end{figure}

There are several ways to motivate $H$. The most direct is to
think of $H$ as a phenomenological quantization of the molecular
vibration. For example, in the case of molecular trimers the two
heavy modes are the antisymmetric stretching and bending of the
molecule, see fig \ref{modes}.

Alternatively, one can start with the Schr\"odinger equation for a
molecule, which indeed has the form of Eq.~(\ref{generalBO}) where
$H_e(x)$ includes the  Coulumb potential of the nuclei and the
electrons and the electronic kinetic energy. Often, and this is
the case in molecules, $H$, is invariant under Gallilean
translation and rigid rotation. The symmetry gives three quantum
numbers $\{P,J,m\}$ and  the spectral analysis of $H_{P,J,m}$ is
now restricted to the ``internal'' nuclear coordinates. In
\cite{mead} one can find a detailed description of this procedure
for a triatomic molecule. $H_{P,J,m}$ has a more complicated
expression than $H$: Fixing the center of mass at the origin and
restricting to an angular momentum subspace replaces the kinetic
energy $-\mu\,\Delta$ by a more general quadratic function of the
of the momenta. However, locally near the crossing this expression
reduces to \cite{mead} Eq.~(\ref{generalBO}).

We shall assume that $H_e(x)$ has discrete spectrum and has smooth
dependence on the coordinates\footnote{Both assumptions are not
realistic; The electronic Hamiltonian has continuous spectrum at
high energy, and because of Coulombic singularities there is no
smoothness in the x dependence. Fortunately, both problems are, by
now, well understood and may be viewed as a technical complication
that, for our purposes, can be left out.}. In addition, since we
shall use the Born Oppenheimer theory as a calculational tool,
rather than a tool of spectral analysis, we shall assume that the
problem has benign qualitative spectral features. For example, we
shall assume that the spectrum of $H$ in the energy range of
interest is discrete, and that the associated wave functions are
localized in space in the classically allowed region, and that
this region is connected. Subtleties associated with tunneling and
other exponentially small phenomena will not concern us here.

Since  the small parameter $\mu$ multiplies the leading derivative
in the $x$ variable, the Born Oppenheimer problem is a version of
semiclassical problem where the operator $H_e(x)$ replaces the
scalar potential $V(x)$.

In molecules there is a second small parameter, $1/c \approx
1/137$, which governs relativistic effects. In particular,
spin-orbit interactions, are of order $1/c^2$. The lowest order of
Born Oppenheimer theory gives an energy scale of $\sqrt\mu$ which
is comparable to $1/c$ but is much larger than $1/c^2$. It is
therefore consistent when discussing Born Oppenheimer to leading
order to disregard spin-orbit. This is also the reason why we
shall not go beyond the leading order.

\subsection{Partial Diagonalization}
The starting point of the Born Oppenheimer method is to consider
$H$ in the basis that diagonalizes the fast (electronic) degrees
of freedom.

 We assume that the electronic Hamiltonian is real, which
is the case in the absence of external magnetic fields. Let $O(x)$
be the orthogonal transformation that diagonalizes $H_e(x)$. If
$H_e(x)$ has simple (non-degenerate) spectrum in the vicinity of
$x$ then $O(x)$ is uniquely determined up to multiplication by a
diagonal matrix with $\pm 1$ on the diagonal. Locally, one can
choose $O(x)$ so that it inherits the smoothness properties of
$H_e(x)$ \cite{kato}. It follows that in the basis that
diagonalizes $H_e(x)$, Eq.~(\ref{generalBO}) takes the form
\begin{equation}\label{OHO}
O^\dag H O =\mu (-i\nabla_x-\,A(x))^2+ E(x),
\end{equation}
 where $E(x)$ is a diagonal matrix whose entries $E_j(x)$ are the
 electronic energy surfaces and
$A(x)=iO^\dag (x)\nabla O(x)$ is a (matrix) gauge field. Since
$O(x)$ is real, $A$ is {\em antisymmetric} and hermitian. $A$ is
responsible to the coupling between electronic energy levels.

 One can  associate
to a crossing point $n$ indices, $\{i_1,\dots,i_n\}$, where
$n=dim\ H_e$ and $i_j=\pm 1$. Let us take  a closed curve around a
crossing point. After such a cycle $O(x)$ must return to itself up
to up to multiplication by a diagonal matrix with entries $\pm 1$
on the diagonal. These entries are the indices of the crossing. It
is known \cite{lh} that for a conic crossing between the $j-th$
and $j+1$ eigenvalues of $H_e(x)$ give $i_j=i_{j+1}=-1$ and all
other indices are, of course, $+1$.

For conic (linear) crossing $O(x)$ flips signs on a circle of
radius $|x|$ \cite{lh}. Therefore its gradient must be of order
$1/|x|$. This makes $A$ of order $1/|x|$. It follows that the
coupling between electronic states diverges like a simple pole
near crossing.

\begin{figure}[htb]
  \centering
  \includegraphics{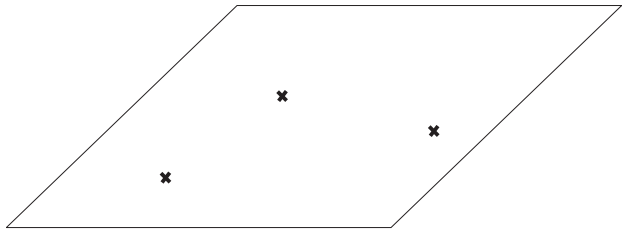}
  \caption{\small With points of crossing removed
the plane becomes multiply connected.}\label{points}
\end{figure}

This completes the local description of the theory. The problem
also has an interesting global aspect. Since $H_e(x)$ is a real
symmetric matrix, the Wigner von Neumann crossing rule \cite{wvn}
says that $H_e(x),\ x\in \mathbb{R}^2$ has, generically, isolated
crossing points. As a consequence, with points of crossing being
removed the plane becomes multiply connected (see fig.
\ref{points})

We can now describe the boundary conditions associated with
Eq.~(\ref{OHO}). The general case of several crossing points can
be complicated but in the case of at most one point of crossing
the situation is simple. In that case, cut the plane from the
crossing point to infinity. On the cut plane $O(x)$ is uniquely
defined in a continuous way. Then the boundary condition on the
j-th component of the wave function associated with
Eq.~(\ref{OHO}) is periodic or anti-periodic according to the
index $i_j$.

\subsubsection{Born Oppenheimer theory near a non-degenerate minimum}
\begin{figure}[htb]
  \centering
  \includegraphics{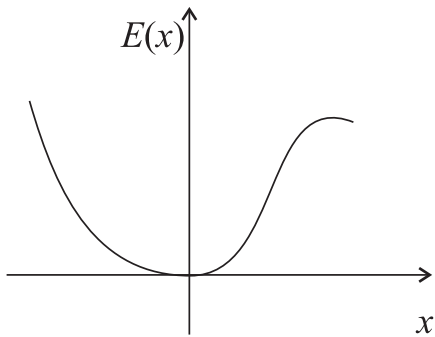}
  \caption{\small A potential energy surface near a non-degenerate minimum.}\label{minimum}
\end{figure}

 Let $x=0$ be a minimizer of an electronic energy surface. Pick
the origin so that the minimum is at zero energy (see Fig.
\ref{minimum}).
 Upon scaling, $x=\mu^{1/4}\xi$, the Born Oppenheimer operator assumes the
form
\begin{equation}\label{UHU2}
\sqrt\mu\left((-i\nabla_\xi-\varepsilon\,A(\varepsilon
\xi))^2+\,\frac{1}{\sqrt\mu}\,E(\varepsilon \xi)\right),\quad
\varepsilon=\mu^{1/4}.
\end{equation}
The (scaled, matrix) potential energy is $O(1)$ for the electronic
energy surface near the minimum, and has gaps of order
$\mu^{-1/2}$ to (scaled) ``excited electronic states''. Suppose
first that there is no crossing. Then the coupling between
electronic levels is small, $ \varepsilon\,A(\varepsilon
\xi)=O(\mu^{1/4})$, and a perturbation argument shows that the
effect on eigenvalues is of order $O(\mu)$ (in unscaled energy)
and of order $O(\mu^{3/4})$ for the wave function. It follows that
the spectral analysis in an energy interval of order $o(1)$ near
the minimum, reduces to an ordinary Schr\"odinger equation (not
matrix valued) with no vector potential (since $A$ vanishes on the
diagonal). This accounts for $o(1/\mu)$ eigenvalues near the
minimum in two dimensions.

\begin{figure}[htb]
  \centering
  \includegraphics{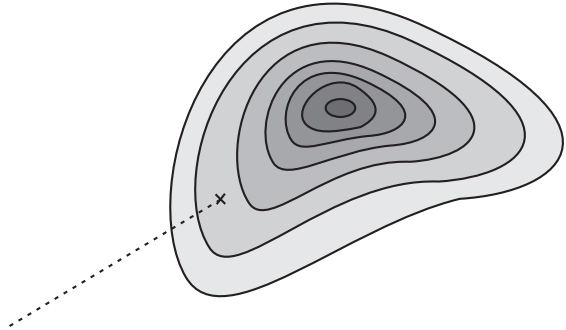}
  \caption{\small For a non-degenerate minimum, when the crossing falls in a
  classically forbidden region, the cut
to infinity can be pushed to the classically forbidden zone, and
one can forget about the crosing to leading order.
}\label{forbidden}
\end{figure}

Now, if there is a point of crossing, there are two possibilities.
The first, and simplest, is that the crossing lies in the
classically forbidden region, see Fig \ref{forbidden}.
Then the divergent vector potential is harmless, since the wave
function is $\exp(-\frac 1 {\sqrt\mu})$ near the crossing. The cut
to infinity can be pushed to the classically forbidden zone, so
the difference between periodic or anti-periodic boundary
conditions is exponentially small, and one can forget about the
crossing altogether. If the crossing point lies in the classically
allowed region, it couples two nuclear Scr\"odinger equations and
ruins the traditional Born Oppenheimer approximation.

\subsubsection{Born Oppenheimer theory near a  degenerate minimizer}
\begin{figure}[htb]
  \centering
  \includegraphics{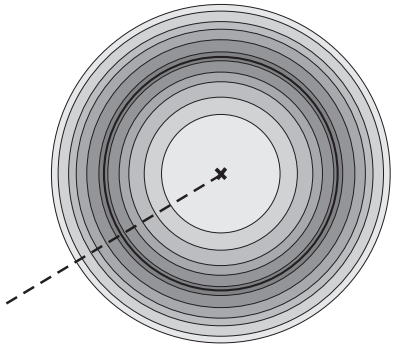}
  \caption{\small A degenerate minimum around a crossing. The solid circle
  represents $\gamma$. Although the crossing falls
  in the classically forbidden zone, it should be taken into account.}\label{encircle}
\end{figure}

It can happen that the crossing must be taken into account
although it lies at the classically forbidden zone. This happens
when the cut can not be pushed to the classically forbidden zone.
This is the case, for example, when the curve $\gamma\in\real^2$
is a minimizer of an electronic energy surface with zero energy
\footnote{Similar arguments apply if the energy associated with
$\gamma$ is sufficiently small}  on $\gamma$. If $\gamma$
encircles a crossing (see Fig. \ref{encircle})
then the cut necessarily intersects $\gamma$ and since $\gamma$
lies in the classically accessible region, the wave function is
large there and it matters if one imposes periodic or
anti-periodic boundary conditions.

\section{Born Oppenheimer theory near crossing}
Consider now the spectral problem near crossing energy of  two
 electronic energy surfaces. Let us set the crossing energy
at $0$ and assume that the crossing is conic. This is the generic
situation.

Upon scaling,  $x=\epsilon\xi,\ \epsilon=\mu^{1/3}$, the Born
Oppenheimer operator assumes the form
\begin{equation}\label{uhu}
\epsilon\left((-i\nabla_\xi-\epsilon\,A(\epsilon
\xi))^2+\,\frac{1}{\epsilon}\,E(\epsilon \xi)\right),\quad
\epsilon=\mu^{1/3}.
\end{equation}
The scaling increases the gaps between the electronic energy
surfaces and decreases the coupling {\em of the crossing pair} to
other levels since $\epsilon\,A_{ij}(\epsilon \xi)=O(\mu^{1/3})$
when $i$ belongs to the pair and $j$ to other levels. On the other
hand,  the crossing pair remains coupled, because the Coulombic
singularity near the crossing says that $\epsilon
A_{i,i'}(\epsilon \xi)=O\left(\frac {1}{|\xi|}\right)$, which is
large when $\xi$ is small. We see that a spectral problem in an
interval of order $o(1)$ near the crossing energy, reduces to a
problem where $H_e(x)$ is a $2\times 2$ matrix up to an error of
order $\mu$ in the eigenvalues and of order $\mu^{2/3}$ in the
eigenfunctions.

 Our aim, in this work, is to describe the wave functions
for states, located at an interval of width that is much smaller
than  $ \mu^{1/3}$ near the crossing. Even though this is a small
interval, it has lots of eigenvalues:
 By Weyl's rule there are many states,
 of order $\mu^{-2/3}$, in an interval of
width $\mu^{1/3}$, in two dimensions. One can expect to find many
states in the interval in question.

Close to the crossing $H_e(x)$ can be expanded in terms of the
Pauli matrices and the unit matrix.  We assume that asymptotically
close to the crossing point $H_e(x)$ is isotropic and
conic\footnote{Reality, isotropy and linearity would allow for an
additional overall scale factor in $H_e$. This scale can be
absorbed in a redefinition of $\mu$.}:
\begin{equation}\label{toy He}
H_e( x)=\sigma\cdot x+ O(x^2).
\end{equation}
 We choose $\sigma_{1,2}$ real\footnote{This
is unconventional, but convenient.}:
\begin{equation}
\sigma_1= \pmatrix{
  0 & 1 \cr
  1 & 0},
\quad \sigma_2=\pmatrix{1 & 0 \cr
  0 & -1}.
\end{equation}$O(x^2)$ accounts for the behaviour far from the
crossing which is not universal, or isotropic.

How restrictive is the assumption that near the crossing $H_e(x)$
is isotropic? Molecules are never isotropic, although some may be
approximately so. Nevertheless, isotropic conic intersections are,
in fact, common. This is a known phenomenon in group theory:
Discrete symmetry can force full continuous symmetry on tensors of
finite rank \cite{herman}. For example, as shown by \cite{mead2},
the
 $D_3$ symmetry of trimers forces isotropy of the conics.

 It follows from the above that if one is interested in the
 local behavior of eigenfunctions for eigenvalues that
 lie in an {\em energy range} that is small compared to $\mu^{1/3}$
and in a {\em spatial neighborhood} of the crossing, $|x|<<1$, the
eigenfunctions satisfy, to leading order in $\mu$, a {\em
canonical system of partial differential equations},
\begin{equation}\label{limit-partial}
\Big(-\mu\,\Delta_{xx} +\sigma\cdot
x\Big)\Psi=\mu^{1/3}\Big(-\,\Delta_{\xi\xi} +\sigma\cdot
\xi\Big)\Psi=0.
\end{equation}
The rotational symmetry allows us to reduce the system
Eq.~(\ref{limit-partial}), to a system of linear, {\em ordinary
differential equation} parameterized by angular momentum $m\in
\mathbb{Z}+\frac 1 2$ (see section \ref{reduced}):
\begin{equation}\label{radial}
\left\{-\frac {d^2}{d \rho^2}-\frac{1}{\rho}\frac{d}{d
\rho}+\frac{m^2+1/4}{\rho^2}+\pmatrix{-\frac{ m}{\rho^2} & \rho
\cr \rho & \frac{ m}{\rho^2}} \right\}{\cal F}_m= 0,\quad
\rho=|\xi |.
\end{equation}
For fixed $m$ the space of solutions of the  ODE is four
dimensional. We shall see that  there is a {\em one dimensional
subspace of solutions} that is well behaved near the origin
$\rho=0$, and near infinity.  As we shall explain, functions in
this space describe the asymptotic behavior of eigenfunctions with
energies near the crossing and spatially close to it. The explicit
expression for these solutions, is described below.

\subsection{The main result}
 We now describe our main result:

{\thm \label{main1} For $m\in \mathbb{Z}+\frac 1 2$, and $m\ge
\frac 1 2$, let
\begin{eqnarray}
{\cal F}_m(\rho)&=&\frac{3^{-\frac 1 2} 6^{-(\frac 5 6 +
\frac{2m}{3}) }\rho^{m-\frac 1 2}} {\Gamma(\frac 2 3)\Gamma(\frac
1 2+\frac m 3 )\,\Gamma(\frac 7 6+\frac m 3)\, } \, \pmatrix{
  { _0F_3}(
  ;\frac{1}{3},\frac{1}{2}+\frac{m}{3},\frac{5}{6}+
  \frac{m}{3};\frac{\rho^6}{6^4})\cr
  \frac{\rho^3}{6+4m}{ _0F_3}(
  ;\frac{4}{3},\frac{3}{2}+\frac{m}{3},\frac{5}{6}+
  \frac{m}{3};\frac{\rho^6}{6^4})}
\cr \cr \cr &-&\frac{3^{-\frac 1 2}\, \,6^{-(\frac 1 6 +
\frac{2m}{3})}\,\rho^{m+\frac 1 2}}{{\Gamma(\frac 1 3)\Gamma(\frac
1 2+\frac m 3 )\Gamma(\frac 5 6+\frac m 3)}}\, \pmatrix
  {
  \frac{\rho^3}{12+8m}{ _0F_3}(
  ;\frac{5}{3},\frac{3}{2}+\frac{m}{3},\frac{7}{6}+
  \frac{m}{3};\frac{\rho^6}{6^4})\cr { _0F_3}(
  ;\frac{2}{3},\frac{1}{2}+\frac{m}{3},\frac{7}{6}+
  \frac{m}{3};\frac{\rho^6}{6^4})}\nonumber
\end{eqnarray}
then \begin{itemize} \item
  ${\cal F}_m(\rho)$, is a solution of the system of differential
 equations, Eq.~(\ref{radial}).
 \item  For small
$\rho$
\begin{equation}\label{as0}
{\cal F}_m(\rho) \to \frac{ \,3^{-\frac 1 2}\rho^{m-\frac 1 2}
}{\Gamma(\frac 1 2 + \frac m 3) 6^{\frac{4m+1}{6}}}
\pmatrix{\frac{1}{\Gamma(\frac 2 3)\Gamma(\frac 7 6 + \frac m 3)
6^{\frac 2 3}} \cr -\frac{\rho}{\Gamma(\frac 1 3)\Gamma(\frac 5 6
+ \frac m 3) }}.
\end{equation}
In particular,  ${\cal F}_{\frac 1 2}(0)\neq 0.$
\item
For large $\rho$
\begin{equation}\label{f_c}
{\cal F}_m(\rho) \to
\frac{1}{(2\pi)^{3/2}\rho^{3/4}}\cos\left(\frac 2 3
\rho^{3/2}-\pi\left(\frac m 3+\frac 1 4\right) \right)\pmatrix{1
\cr -1}
\end{equation}
\end{itemize}
}
\begin{figure}[htb]
  \centering
  \includegraphics{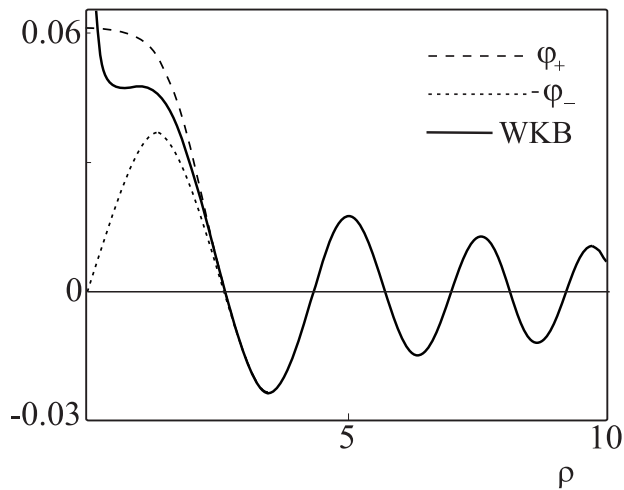}
  \caption{\small The components of ${\cal F}_m(\rho)$ for $m=\frac 1 2$
and their asymptotic form given by (\ref{f_c}). The function is
compared with the WKB Born Oppenheimer wave function.}\label{f1}
\end{figure}
{\thm \label{main2} For $m\in \mathbb{Z}+\frac 1 2$, and $m\ge
\frac 1 2$, let \begin{equation} \Psi_m(r,\theta)= \mu^{-1/4}
e^{im\theta}\,\pmatrix{
  e^{i\theta/2} & i e^{-i\theta/2} \cr
  i e^{i\theta/2} &  e^{-i\theta/2}
}\,{\cal F}_m(\mu^{-1/3}r)
\end{equation}
then \begin{itemize} \item $\Psi_m$ is a solution of the system of
PDE, Eq.~(\ref{limit-partial}) where $x=(r\cos\theta,r\sin\theta
)$.
\item The m-th component of an eigenfunction of Eq.~(\ref{generalBO})
near crossing, i.e. with eigenvalue $|E|<<\mu^{1/3}$ and for
$x=(r\cos\theta,r\sin\theta),\ r<<1$ is, to leading order,
proportional to $\Psi_m(r,\theta)$.
\item The amplitude of $\Psi_m$ is independent of $\mu$ in the
region $r >> \mu^{1/3}$.
\item Near the crossing, $r\approx \mu^{1/3}$, the amplitude of
the wave function $\Psi_m = O(\mu^{-1/4})$
\end{itemize}
 } {\remark The most interesting aspect of the solution is that  the
 wave function has {\em large amplitude} at the
crossing region in the limit of small $\mu$. As we shall see, this
result follows from arguments that do not rely on the explicit
form of the solution, but do depend on the fact that in the
non-mixing region, $r>>\mu^{1/3}$ the solution has a WKB form in
the radial direction. }

{\remark The function $\Psi_m$ {\em does not} describe the
behavior of wave functions in the far zone where $r >1$. The
behaviour in the far zone is not universal and depends on the
details of the electronic energy surface . The far zone is
described by standard Born Oppenheimer, so $\Psi_m$ gives
complementary information.}

{\remark There are  $o\left(\mu^{-2/3}\right)$ states in the
relevant energy interval $o\left(\mu^{1/3}\right)$ near the
crossing disregarding $m$. For a given $m$  there are only
$\mu^{-1/6}$ levels in this interval. This says that near the
crossing $m$ is bounded by order $\mu^{-1/2}$.}

{\remark We shall actually only prove the theorem in the special
case that $H_e(x)$ is rotationally symmetric. The symmetry
decouples channels with different angular momenta. We believe that
mixing of angular momenta in the far zone is only a technical
complication and that the result also holds without rotational
symmetry in the far zone. }
\section{The Rotationally Symmetric Case}
 In the following we describe a derivation of the main result for
  a Born Oppenheimer  model that is rotationally
symmetric. With rotational symmetry we can reduce the spectral
problem of a PDE to a spectral problem of an ODE.  No real
molecule is rotationally symmetric and the general case leads to
mixing of $m$ channels. We believe that this is only a technical
complication.

We  require invariance of $H_e(x)$ under infinitesimal rotations
 in the nuclear and electronic
Hilbert spaces. Such a rotation is generated by
\begin{equation}\label{jz}
J_3=L_3+\frac 1 2 \sigma_3=-i x_1 \frac{\partial}{\partial x_2} +
i x_2 \frac{\partial}{\partial x_1} +\frac 1 2 \sigma_3,
\end{equation}
with $$ \sigma_3=\pmatrix{0 & i \cr -i & 0}.$$ The $L_3$ part
generates $SO(2)$ rotations in the nuclear Hilbert space
($x_1-x_2$ plane), whereas the $\frac 1 2 \sigma_3$ part generates
a rotation in the electronic Hilbert space. $J_3$ does not have
the meaning of total angular momentum since the Pauli matrices do
not represent spin. Isotropy means that $J_3$ commutes with
$H_e(x)$:
\begin{eqnarray}
0=[J_3,H_e(x)] \nonumber
\end{eqnarray}
 The most general form of $H_e(x)$ for a two level
system that is rotationally symmetric and real is:
\begin{equation}\label{rot_V}
H_e(x)=Q_0(r)+Q_1(r)(x\cdot \sigma)+ Q_2(r)(x\times\sigma),\quad
r=|x|
\end{equation}
An additional $Q_3(r) \sigma_3$ term in (\ref{rot_V}) is allowed
by rotational invariance but it is forbidden by time reversal
symmetry, since $\sigma_3$ is imaginary.

\begin{figure}[htb]
  \centering
  \includegraphics[height=4.cm]{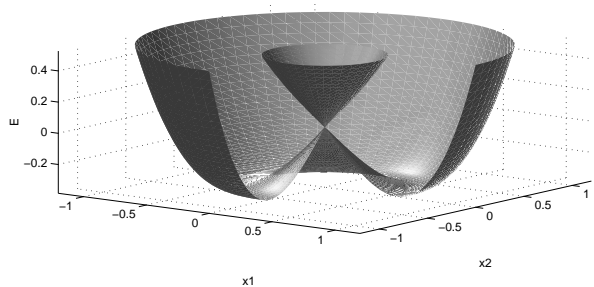}
  \caption{\small The electronic energy surfaces of $H_e(x)$}\label{fig3D}
\end{figure}

 The energy surfaces of
$H_e(x)$ are equal to
\begin{equation}\label{surfaces}
E_\pm(r)=Q_0(r) \pm q (r);\quad q (r) = r\sqrt{Q_1^2(r)+Q_2^2(r)}
\end{equation}
We shall assume that $Q_{0,1,2}$ are smooth functions of $r$ and
that $Q_1(0)=1$ while\footnote{$Q_2(0)$ can be always set to zero
by an appropriate rotation in the $x$ space only, i.~e.~by an
appropriate choice of the heavy coordinates.} $Q_{0,2}(r)=O(r^2)$
for small $r$. This gives conic intersection at zero with
$E_\pm(r)=\pm r$, see Fig.~\ref{fig3D}.

\subsection{The radial Hamiltonian}\label{reduced}
The spectral subspace of $J_3$, with eigenvalue $m$ is spanned by
\begin{equation} e_1=
e^{i(m+1/2)\theta} \pmatrix{1\cr i}\, \quad e_2=
e^{i(m-1/2)\theta} \pmatrix{i\cr 1}
\end{equation}
 $m$ must be half odd integer for $e_{1,2}$
to be univalued, i.e. $m\in\mathbb{Z}+\frac 1 2$.

Since
\begin{eqnarray}
-\Delta\, e_1= \frac{(m+\frac {1}
{2})^2}{r^2}\,e_1,&&\quad-\Delta\, e_2= \frac{(m-\frac {1}
{2})^2}{r^2}\,e_2, \cr  x \cdot \sigma \, e_1= r\,e_2,&&\quad
  x \cdot \sigma \,e_2=r\,e_1, \cr
  x \times \sigma\,e_1= -ir\,e_2,&& \quad
   x \times \sigma\,e_2= ir\,e_1,
\end{eqnarray}
in terms of the basis $\{e_1, e_2\}$ we obtain for the radial
equation:
\begin{equation}\label{radial1}
 H(m,\mu)=- \mu\left(\frac {d^2}{d r^2}+\frac{ 1}{r}\frac{d}{d r}-\frac{ 1}{4 r^2}\right)
 + H_e(r,\mu,m)
\end{equation}
with
\begin{equation}\label{vr}
 H_e(r,\mu,m)=Q_0(r)+rQ_1(r)\sigma_1+r Q_2(r)\sigma_3 -
 \frac{\mu}{r^2}\left({ m}\sigma_2 - m^2\right)
\end{equation}
Scaling $r=\mu^{1/3}\rho$ we get Eq.~(\ref{radial}), to leading
order in $\mu$, for $0\le \rho << \mu^{-1/3}$.

{\remark The radial Hamiltonian, $H_e(m,\mu,r)$, actually has no
level crossing. This, by itself, does not ameliorate the mixing of
the two levels for now the gap in the spectrum of $H_e(m,\mu,r)$
is of order $\mu^{1/3}$. The smallness of this gap leads to mixing
of the electronic levels.}

 We shall restrict ourselves to $m>0$. Since
 \begin{equation}
 \sigma_1\, H^*_e(r,\mu,m)\sigma_1= H_e(r,\mu,-m),\end{equation}
   $H(m,\mu)$ and
$H(-m,\mu)$ are isospectral and the radial part of the function
with $-m$ can be obtained from the one with $+m$ by interchanging
upper and lower components and taking complex conjugates.

\subsection{The Indicial equation} The origin, $\rho =0$ is
regular-singular point \cite{boyce} of the equation. Substituting
\begin{equation}
\pmatrix{\rho^{\alpha} \cr \rho^{\beta}}(1+O(\rho))
\end{equation}
into  (\ref{radial}) we obtain the roots ${\alpha}=\pm(m-\frac 1
2)$ and ${\beta}=\pm(m+\frac 1 2)$. Equation (\ref{radial})
 therefore has four linearly independent solutions, which
asymptotically near the origin, behave like
\begin{equation}\label{asnear0}
\pmatrix{\rho^{m - \frac 1 2} \cr 0};\,\, \pmatrix{\rho^{-m +
\frac 1 2} \cr 0};\,\, \pmatrix{0 \cr \rho^{m + \frac 1 2}};
\,\,\pmatrix {0 \cr \rho^{-m - \frac 1 2}}
\end{equation}
(\ref{asnear0}) is correct only for $|m|\neq\frac 1 2$. The case
 $|m|=\frac 1 2$ requires special treatment, because of the two
degenerate roots in the upper component. For $|m|=\frac 1 2$ the
four linearly independent solutions behave asymptotically near the
origin like \cite{boyce}
\begin{equation}\label{asnear0log}
\pmatrix{1 \cr 0};\quad \pmatrix{\ln(\rho) \cr 0};\quad \pmatrix{0
\cr \rho}; \quad \pmatrix {0 \cr 1/\rho}
\end{equation}
 We see that for any $m$ there
are always two solutions which are bounded near the origin and two
others which are divergent. Since a smooth Hamiltonian can give
rise to smooth eigenfunctions only \cite{simon} in the four
dimensional space of solutions to the differential equation, there
is a two dimensional subspace of admissible solutions, the ones
which are well behaved at the origin.

\subsection{Solution to the ODE}
In this section we show that the solutions of Eq.~(\ref{radial})
that are regular at the origin, can be explicitly constructed in
terms of certain Hypergeometric functions.

 {\thm \label{hyper_solutions} The solutions of
(\ref{radial}) which are bounded at the origin are spanned by:
\begin{eqnarray}\label{hyper_solutions1}
{\cal F}_{m}^{(1)}(\rho)&=& \pmatrix{\varphi_{+ }^{(1)}(\rho) \cr
\varphi_{- }^{(1)}(\rho)}=\pmatrix{
  \rho^{m-\frac{1}{2}}{ _0F_3}(
  ;\frac{1}{3},\frac{1}{2}+\frac{m}{3},\frac{5}{6}+
  \frac{m}{3};\frac{\rho^6}{6^4})\cr \frac{\rho^{m+\frac{5}{2}}}{6+4m}{ _0F_3}(
  ;\frac{4}{3},\frac{3}{2}+\frac{m}{3},\frac{5}{6}+
  \frac{m}{3};\frac{\rho^6}{6^4})};\cr \cr
  {\cal F}_m^{(2)}(\rho)&=& \pmatrix{\varphi_{+ }^{(2)}(\rho) \cr
\varphi_{-}^{(2)}(\rho)}= \pmatrix
  {
  \frac{\rho^{m+\frac{7}{2}}}{12+8m}{ _0F_3}(
  ;\frac{5}{3},\frac{3}{2}+\frac{m}{3},\frac{7}{6}+
  \frac{m}{3};\frac{\rho^6}{6^4})\cr \rho^{m+\frac{1}{2}}{ _0F_3}(
  ;\frac{2}{3},\frac{1}{2}+\frac{m}{3},\frac{7}{6}+
  \frac{m}{3};\frac{\rho^6}{6^4})},\phantom{==}
\end{eqnarray}
where $_0F_3(;a,b,c;x)$ are the generalized hypergeometric
functions of the kind $_0F_3$.}

 \noindent{\bf Proof:}
Under scaling, $\rho\to\lambda\rho$, Eq.~(\ref{radial}) transforms
to
\begin{equation}\label{radial-scaled}
\left\{-\frac {d^2}{d \rho^2}-\frac{1}{\rho}\frac{d}{d
\rho}+\frac{m^2+1/4}{\rho^2}+\pmatrix{-\frac{ m}{\rho^2}
&\lambda^3 \rho \cr \lambda^3\rho & \frac{ m}{\rho^2}}
\right\}{\cal F}_m(\lambda\rho)= 0.
\end{equation}
In particular, the equation is invariant under scaling by
$\lambda$, a cube root of unity, $\lambda=e^{2\pi i/3}$. (Note
that this feature is lost when one considers solutions of the
equation for non-zero eigenvalue.) By an analog of Bloch theorem,
the solution is a product of an eigenfunction of the scaling
transformation and a periodic function under scaling by $e^{2\pi
i/3}$.
 $\rho^\alpha$ is an eigenfunction of
the scaling transformation with eigenvalue $\lambda^\alpha$. Hence
that ${\cal F}_m$ must be of the form $\rho^\alpha \,{\cal
G}(\rho^3)$. The indicial equation fixes $\alpha=m-1/2$. ${\cal
G}_m(\rho^3)$ is then an analytic function of its argument.

The space of solutions regular of this kind is two dimensional and
gives a representation of $D_3$, the group of discrete rotations
by $2\pi/3$. Since the only complex irreducible representations
\cite{wigner} of $D_3$ are the complex numbers $\omega$, such that
$\omega ^ 3 = 1$, one can always find a basis ${\cal
F}_{m}^{(1)}(\rho), {\cal F}_m^{(2)}(\rho)$ such that $${\cal
F}_{m}^{(j)}(e^{2\pi i/3}\rho)=\omega_j{\cal F}_{m}^{(j)}(\rho).$$
This condition fixes the solutions in Eq.~(\ref{hyper_solutions1})
where $\omega_1=e^{2\pi i (m-1/2)/3}$ and $\omega_2=e^{2\pi i
(m+1/2)/3}$.

  To relate ${\cal G}_m$ to hypergeometric functions we
turn the two coupled second order equations (\ref{radial}) into a
scalar fourth order equation for each component. The equations
obtained for the component  $\varphi_+$ and $\varphi_-$ can be
written, with $\zeta=\rho^6/6^4$, and ${\rm
D}=\zeta\frac{d}{d\zeta}$, in the form:
\begin{eqnarray}\label{4order}
\left\{{\rm D}\left({\rm D}-\frac{2}{3}\right) \left({\rm
D}+\frac{m}{3}-\frac{1}{2}\right) \left({\rm
D}+\frac{m}{3}-\frac{1}{6}\right)-\zeta\right\}
\zeta^{\frac{-m+1/2}{6}}\varphi_+(\zeta)=0 \cr \left\{{\rm
D}\left({\rm D}-\frac{1}{3}\right) \left({\rm
D}+\frac{m}{3}-\frac{1}{2}\right) \left({\rm
D}+\frac{m}{3}+\frac{1}{6}\right)-\zeta\right\}
\zeta^{\frac{-m-1/2}{6}}\varphi_-(\zeta)=0
\end{eqnarray}
The generalized hypergeometric function $_0F_3(;a,b,c;\zeta)$ is
defined by \cite{handbook}:
\begin{equation}
{_0F_3}(;a,b,c;\zeta)= \sum_{k=0}^\infty
\frac{\Gamma(a)\Gamma(b)\Gamma(c)}{k!\Gamma(k+a)\Gamma(k+b)\Gamma(k+c)}
\zeta^k.\end{equation} It is a matter of calculation to see that
it satisfies the differential equation:
\begin{equation}\label{hyp_eq}
\left\{{\rm D}({\rm D}+a-1) ({\rm D}+b-1) ({\rm
D}+c-1)-\zeta\right\} {_0F_3}(;a,b,c;\zeta)=0.
\end{equation}
Eq.~(\ref{4order}) is a special case of this. Note, however, that
we are {\em not} free to pick {\em both}  $\varphi_+^{(1)}$ and
$\varphi_-^{(1)}$ as the hypergeometric functions corresponding to
Eq.~(\ref{hyp_eq}). We can pick one, and then the other is
determined by Eq.~(\ref{radial}).

To get ${\cal F}^{(1)}_m$ we pick the upper component to be the
hypergeometric function that solves Eq.~(\ref{4order}), i.e. $$
\varphi_{+}^{(1)}(\rho)=\rho^{m-\frac{1}{2}}{ _0F_3}\left(
  ;\frac{1}{3},\frac{1}{2}+\frac{m}{3},\frac{5}{6}+
  \frac{m}{3};\frac{\rho^6}{6^4}\right). $$
The lower component,$\varphi_-^{(1)}$, is determined by
(\ref{radial}), which gives us the relation
\begin{equation}\label{phi_down_1_formula}
\varphi_{-}^{( 1)}(\rho)= \frac{\rho^{m-\frac 1
2}}{\rho^3}\left\{\rho\frac{d}{d\rho}\left(\rho\frac{d}{d\rho}+2m-1\right)\right\}
\frac{\varphi_{+}^{(1)}(\rho)}{\rho^{m-\frac 1 2}}
\end{equation}
With the identities \cite{handbook,prudnikov}
\begin{eqnarray}
\left(\zeta\frac{d}{d\zeta}+c-1\right){_0F_3}(;a,b,c;\zeta)&=&(c-1){_0F_3}(;a,b,c-1;\zeta)
\cr \frac{d}{d\zeta} {_0F_3}(;a,b,c;\zeta)&=&\frac{1}{abc}
{_0F_3}(;a+1,b+1,c+1;\zeta) \nonumber
\end{eqnarray}
we obtain \begin{equation} \varphi_{-}^{(
1)}(\rho)=\frac{1}{6+4m}\rho^{m+\frac{5}{2}}{ _0F_3}\left(
  ;\frac{4}{3},\frac{3}{2}+\frac{m}{3},\frac{5}{6}+
  \frac{m}{3};\frac{\rho^6}{6^4}\right).\end{equation}

 The second solution, ${\cal F}^{(2)}_m$, is obtained by picking
 $\varphi_-^{(2)}$ to be the hypergeometric solution to Eq.~(\ref{4order}).
 Namely,  \begin{equation}
\varphi_{-}^{ (2)}(\rho)=\rho^{m+\frac{1}{2}}{ _0F_3}\left(
  ;\frac{2}{3},\frac{1}{2}+\frac{m}{3},\frac{7}{6}+
  \frac{m}{3};\frac{\rho^6}{6^4}\right).
\end{equation} and with a relation $$ \varphi_{+}^{( 2)}(\rho)=
\frac{\rho^{m+\frac 1
2}}{\rho^3}\left\{\rho\frac{d}{d\rho}\left(\rho\frac{d}{d\rho}+2m+1\right)\right\}
\frac{\varphi_-^{(2)}(m;\rho)}{\rho^{m+\frac 1 2}} $$ similar to
(\ref{phi_down_1_formula}) one computes the upper component of the
second solution: $$ \varphi_{+}^{( 2)}(\rho)=\frac
{1}{12+8m}\rho^{m+\frac{7}{2}}{ _0F_3}\left(
  ;\frac{5}{3},\frac{3}{2}+\frac{m}{3},\frac{7}{6}+
  \frac{m}{3};\frac{\rho^6}{6^4}\right)
\hfill \blacksquare$$

\subsection{The well behaved solutions}
We have seen that of the four dimensional family of solution  of
Eq.(\ref{radial}) there is a distinguished two dimensional family
that is well behaved near the origin. We shall now show that there
is a three dimensional family that is well behaved at infinity:
 {\thm \label{final_wkb_thm}
 \begin{itemize}
 \item In the four dimensional space of solutions of
 Eq.~(\ref{radial}) there is a three dimensional family of solutions that
  vanish at infinity, and one dimensional subspace that
 diverges exponentially at infinity.\item The
solutions of (\ref{radial}) for $\rho
>>1$ are (asymptotically) spanned by the four dimensional family :
\begin{eqnarray}\label{final_wkb}
  \rho^{-3/4}\,\exp \left({-\frac 2 3 \rho^{3/2}}\right)\,\pmatrix{1\cr 1};&&\quad
 \rho^{-3/4}\,\exp\left({\frac 2 3 \rho^{3/2}}\right)\,\pmatrix{1\cr 1};\nonumber
 \\
 \rho^{-3/4}\,\cos\left(\frac 2 3 \rho^{3/2}\right)\,\pmatrix{1\cr -1};&&\quad
  \rho^{-3/4}\,\sin\left(\frac 2 3 \rho^{3/2}\right)\,\pmatrix{1\cr -1}.
\end{eqnarray}
\item The exponential blow up of
${\cal F}_m^{(1,2)}$ of Eq.~(\ref{radial}) is given by
\begin{eqnarray}
2{\cal F}_m^{(1)}(\rho)&\to&{\Gamma\left(\frac 1
3\right)\Gamma\left(\frac 1 2+\frac m 3 \right)\Gamma\left(\frac 5
6+\frac m 3\right)6^{\frac 1 6 + \frac{2m}{3}}}
\rho^{-3/4}\,\exp\left({\frac 2 3
\rho^{3/2}}\right)\,\pmatrix{1\cr 1}\nonumber \\ 2 {\cal
F}_m^{(2)}(\rho)&\to& {\Gamma\left(\frac 2
3\right)\Gamma\left(\frac 1 2+\frac m 3 \right)\Gamma\left(\frac 7
6+\frac m 3\right)6^{\frac 5 6 +
\frac{2m}{3}}}\rho^{-3/4}\,\exp\left({\frac 2 3
\rho^{3/2}}\right)\,\pmatrix{1\cr 1}\nonumber
\end{eqnarray}
\item The solution to Eq~(\ref{radial}) that vanishes at the
origin and at infinity is a multiple of
\begin{eqnarray}
{\Gamma\left(\frac 2 3\right)\Gamma\left(\frac 1 2+\frac m 3
\right)\Gamma\left(\frac 7 6+\frac m 3\right)6^{\frac 5 6
}}\,{\cal F}_m^{(1)}(\rho) -\nonumber \\ {\Gamma\left(\frac 1
3\right)\Gamma\left(\frac 1 2+\frac m 3 \right)\Gamma\left(\frac 5
6+\frac m 3\right)6^{\frac 1 6 }}\,{\cal
F}_m^{(2)}(\rho).\nonumber
\end{eqnarray}
\end{itemize}
}
{\bf Proof:} $H_e(\rho,\mu,m)$ for Eq.~(\ref{radial}) is, for
$\rho >>1$,
\begin{equation}H_e(\rho,\mu, m)=
\frac{m^2+1/4}{\rho^2}+\pmatrix{-\frac{ m}{\rho^2} & \rho \cr \rho
& \frac{ m}{\rho^2}}\approx \pmatrix{0 & \rho \cr \rho &0}
\end{equation}
With eigenvalues $\pm\rho$. It follows that the  solution for
large $\rho$ reduces to the study of two uncoupeld equations:
\begin{equation}\label{radial2} \left(-\frac
{d^2}{d \rho^2}-\frac{1}{\rho}\frac{d}{d
\rho}\pm\rho\right)\psi=0.
\end{equation} Since $\rho$ is large, these can be solved by WKB
to give the first part of the theorem. The blow up of the
solutions at infinity can be obtained from the relation (omitting
the exponentially decaying part)
\begin{equation}\label{blow}
  _0 F_3(;a,b,c;x) \to
\frac{\Gamma(a)\Gamma(b)\Gamma(c)}{2(2\pi)^{3/2}}x^\gamma\left(e^{4x^{\frac
1 4}}+2\cos(4x^{\frac 1 4}+2\pi\gamma) \right),
\end{equation}
with $\gamma=-\frac{a+b+c-3/2}{4}$. This relation can be obtained
by studying the asymptotic behavior of the coefficients in the
series of $_0F_3$. Alternatively, in \cite{wimp}, the asymptotic
form of the generalized hypergeometric functions $_pF_q$ is
derived, and the formula given there reduces to (\ref{blow}) after
substituting $q=0, p=3$, computing the summations and omitting the
exponentially decaying part.

>From this the rest follows, as well as the proof of theorem
\ref{main1}.
 \hfill $\blacksquare$

 It remains to explain how eigenvectors are related to these well
 behaved solutions. The point is that the
 canonical differential equation approximates the eigenvalue
 equation only for $r<<1$, or, equivalently, for $\rho
 <<\mu^{-1/3}$. Consider an eigenfunction. Far from the crossing
 this eigenfunction can be approximated by a WKB solution, and it
 is clear that this WKB solution can be approximated by WKB solution of the
 canonical problem in the interval $\mu^{-1/3}>>\rho >>1$. The
 component that blows up must have an exponentially small
 amplitude, of order $\exp (-\frac{2}{3\sqrt\mu}) $, and, to leading
 order can be neglected near the crossing.

\section{Anomalous Mixing}
The basic and fundamental observation of the Born Oppenheimer
theory is the emergence of the energy scale $\sqrt\mu$ in
molecular spectra associated with vibrations. Perhaps the most
interesting observation that results from the analysis of the
Born Oppenheimer theory near crossing, is that emergence of the
   scale, $\mu^{1/6}$, associated with mixing at crossing.
 $\mu^{1/6}$ is normally not a
small number: In molecules $\mu^{1/6}\sim .2$.

Classically, a uniform density on the energy shell,
$\delta(p^2+V(x)-E)$ implies that, in two dimensions, the spatial
density is also uniform on the classically allowed region. The
region where there is substantial mixing between the two
electronic energy surfaces has linear dimensions that scale like
$\mu^{1/3}$. For a crossing point in two dimension the volume
characterizing the mixing therefore scales like $\mu^{2/3}$. The
semiclassical expectation is therefore that mixing near crossing
should scale like the area $\mu^{2/3}$.

For isotropic crossing we found that the wave function has
anomalously {\em large} amplitude in the mixing region for values
of azymuthal quantum numbers that are small compares to
$\mu^{-1/2}$. For these, from theorem \ref{main2}, the amplitude
in the near zone scales like $\mu^{-1/4}$. This implies that the
total mixing weight scales like $\mu^{1/6}$.

It would be interesting to have a more complete picture of the
mixing near non-isotropic  crossing and also for chaotic systems.

 \section*{Acknowledgments}
We thank Richard Askey and Jet Wimp for helpful correspondence about Hypergeometric
functions; J. Ax and S. Kochen for pointing sign errors in a previous version of the
manuscript; C. Alden Mead for helpful suggestions; M.V. Berry for encouraging us to
look for a special function that characterizes the crossing and E. Berg, M. Baer, R.
Englman, A. Elgart and L. Sadun for helpful discussions. This research was supported
in part by the Israel Science Foundation, the Fund for Promotion of Research at the
Technion and the DFG.

\end{document}